# Fermilab Program and Plans

**Dmitri Denisov**

*Fermi National Accelerator Laboratory,*
*PO Box 500, Batavia IL 60510, USA*
*E-mail*: denisovd@fnal.gov

ABSTRACT: This article is a short summary of the talk presented at 2014 Instrumentation Conference in Novosibirsk about Fermilab's experimental program and future plans. It includes brief description of the P5 long term planning progressing in US as well as discussion of the future accelerators considered at Fermilab.

# Contents

## 1. Planning Long Term Scientific Program in US

In US high energy physics process of developing long term scientific program consists of the following five main steps:

- Groups of scientists develop proposals for future projects/experiments.

- "Snowmass" community wide process discusses proposals, evaluates physics reach and costs and summarizes outcome in a written form.

  - Organized by Division of Particles and Fields (DPF) – professional organization, not Laboratories or NSF (National Science Foundation) or DOE (Department of Energy).

- P5 committee (Particle Physics Projects Prioritization Panel) is formed consisting of ~25 scientists representing all areas of particle physics.

  - The committee recommends priorities for funding based on available funds and expected cost of the projects.

- HEPAP (High Energy Physics Advisory Panel) appointed by DOE reviews the P5 proposal and recommends it to be considered by DOE/NSF.

- DOE/NSF fund recommended projects based on the available funds.

The process described is initiated about every five years and most recent was initiated in 2013. The DPF charge for "Snowmass 2013" reads: "To develop the community's long term physics aspirations. Its narrative will communicate the opportunities for discovery in high energy physics to the broader scientific community and to the Government".

Snowmass process was organized around so called "frontiers": energy, intensity, cosmic, instrumentation (which is the subject of Novosibirsk conference), facilities (mainly new accelerators), education and outreach, and theory. Time scale for the proposals to be discussed is ~10 years, taking into account ~20 years time span. The process continued for about a year and culminated in ~10 days community meeting at the University of Minnesota late July 2013 where in depth discussion of all proposals, their scientific merit, facilities required as well as cost have been presented. The outcome of the Snowmass process is a large number of documents which are summarized on the Snowmass Web page [1].

## 1.1 Big Questions for Particle Physics

One of the main outcomes of the Snowmass process was the development of "big questions" for particle physics. They are

- How do we understand the Higgs boson? Why does it condense and acquire a vacuum value throughout the universe? Is there one Higgs particle or many? Is the Higgs particle elementary or composite?
- What principle determines the masses and mixings of quarks and leptons? Why is the mixing pattern apparently different for quarks and leptons? Why is the CKM CP phase nonzero? Is there CP violation in the lepton sector?
- Why are neutrinos so light compared to other matter particles? Are neutrinos their own antiparticles? Are their small masses connected to the presence of a very high mass scale? Are there new interactions invisible except through their role in neutrino physics?
- What mechanism produced the excess of matter over antimatter that we see in the universe? Why are the interactions of particles and antiparticles not exactly mirror opposites?
- Dark matter is the dominant component of mass in the universe. What is the dark matter made of? Is it composed of one type of new particle or several? Are the dark matter particles connected to the particles of the Standard Model, or are they part of an entirely new dark sector of particles?
- What is dark energy? Is it a static energy per unit volume of the vacuum, or is it dynamical and evolving with the universe? What principle determines its value?
- What did the universe look like in its earliest moments, and how did it evolve to contain the structures we observe today?

Definition of what are the main topics for experimental and theoretical studies is an important first step to select most interesting and scientifically important projects for the next decade.

## 2. Fermilab's Scientific Program

### 2.1 Tevatron

The Tevatron accelerator complex presented in Fig. 1 was the center of the Laboratory scientific activities for almost 25 years. Two general-purpose collider detectors CDF and DZero published over 1000 papers cementing the Standard Models. Discovery of the top quark in 1995 was fundamental discovery as well as evidence for the Higgs boson production and decay to a pair of b-quarks in 2012. On the first day of Novosibirsk conference CDF and DZero jointly announced another discovery: observation of t-channel single top quark production with 6.3 sigma significance. More results, especially in very high precision measurements, are expected from 12 fb$^{-1}$ data set of proton-antiproton collisions at 2 TeV delivered by the Tevatron to both collider experiments.

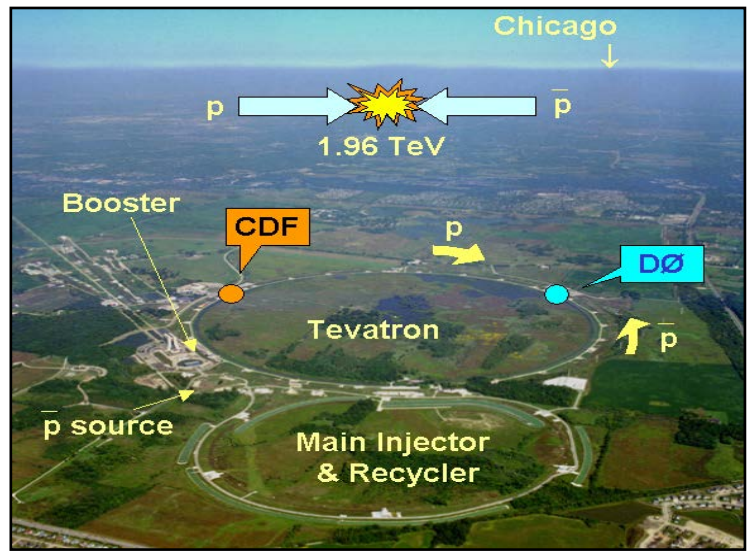

Fig. 1 Tevatron accelerator complex and two detectors CDF and DZero.

## 2.2 Fermilab Accelerator Complex

With Tevatron shutdown late 2011 the Fermilab accelerator complex is now concentrating on delivering high intensity proton beams at up to 150 GeV to a wide range of programs including neutrino experiments, high intensity muon beam experiments as well as test beams. The scheme of the current Fermilab's accelerator complex is presented I n Fig. 2.

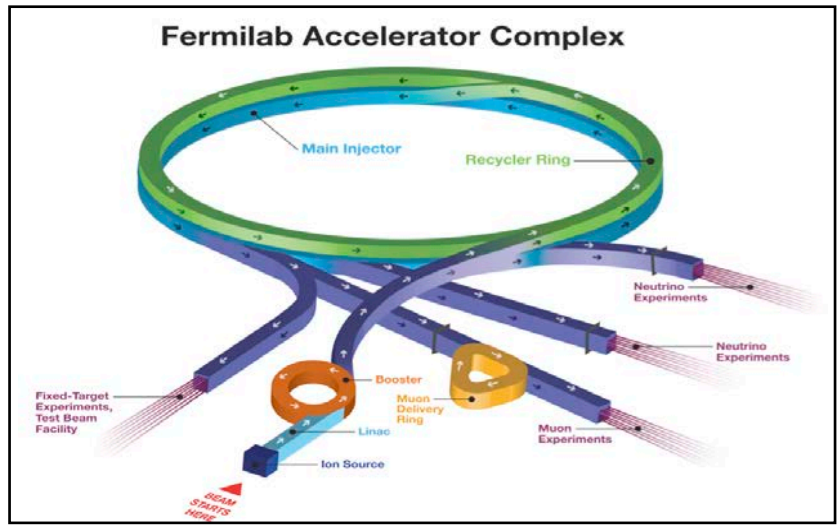

Fig. 2. Current Fermilab's accelerator complex.

In the near term Fermilab is planning to construct a multi-MW proton linear accelerator with flexible beam structure based on SCRF technology. Power will reach 1 MW at 1 GeV and more at 3-8 GeV with over 2 MW of power to neutrino program at 120 GeV. Such upgrade could serve multiple experiments over a broad energy range and serve as a platform for future neutrino and muon facilities (including muon collider).

## 2.3 Neutrino Program

Neutrinos remian one of the most puzzling particles of the Standard Model. Studies of neutrino properties, including mixing matrix parameters, could shed light on many interesting topics, including matter-antimatter asymmetry of the Universe.

The MINOS experiment is a long-baseline neutrino oscillation experiment operating at Fermilab with two detectors, near and far in order to improve beam flux calculations and reduce neutrino interaction uncertainties. The L/E ratio for the MINOS experiment is ~500 km/GeV. Figure 3 presents one of the recent MINOS results which is indicative that the experiment is starting to approach sensitivity to the mass hierarchy and CP violation in neutrino mixing. MINOS was among experiments which rejected claims that neutrinos travel faster than the speed of light.

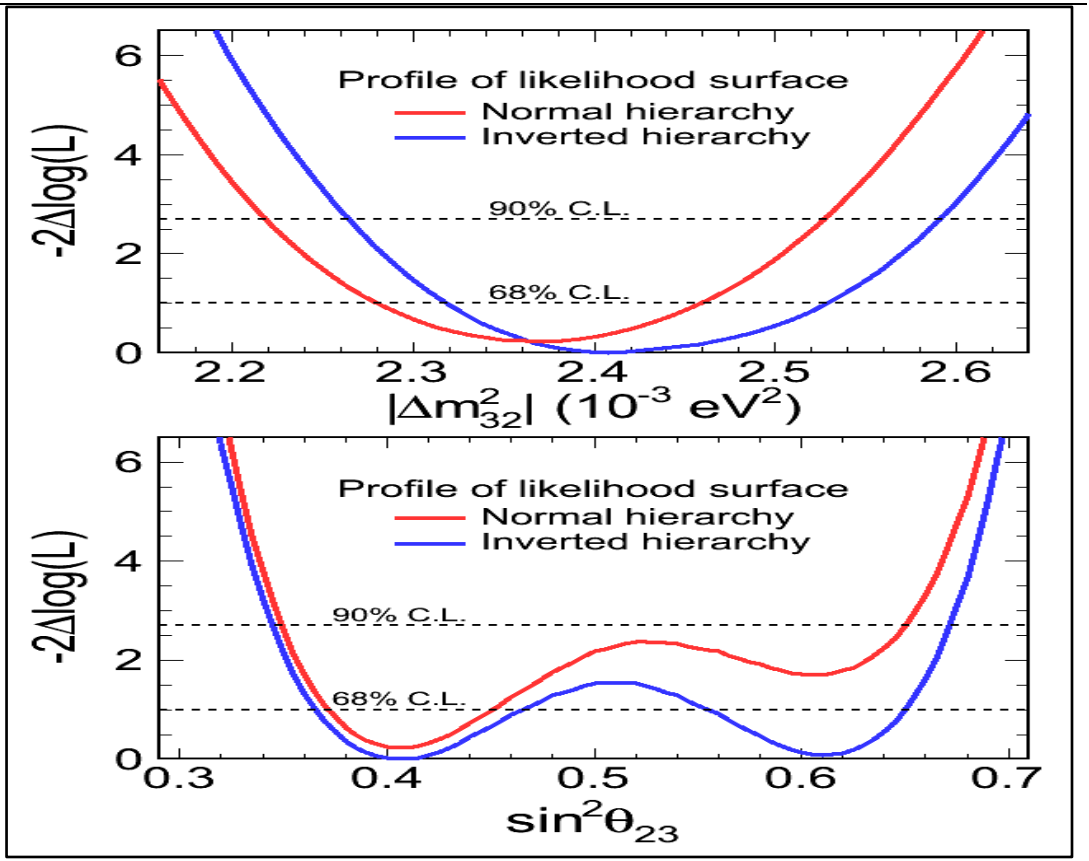

Fig. 3. Recent MINOS results are becoming sensitive to mass hierarchy and CP violation.

The MINERvA experiment is located just in front of MINOS in the neutrino beam with energies of a few GeV and is studying neutrino interactions in unprecedented detail on a variety of different nuclei – He, C, $CH_2$, $H_2O$, Fe, Pb. This experiment provides important information for all neutrino based experiments.

MicroBooNe is a new neutrino experiment with 170 ton of LAr based on time projection technology with a few meters of electrons drifts in LAr. This is a new technology development which could potentially improve neutrino detection substantially. This experiment is expected to start taking data in late 2014.

NOvA is yet another brand new neutrino "off axis" experiment to start data collection in 2014. The main idea of this experiment is to use an "off axis" beam of neutrino which has rather small energy spread with a peak at ~1 GeV. The detector total weight is 15 kton and it is located on surface close to the MINOS detector and uses liquid scintillator and WLS fibers with APD as readout.

Fermilab's future neutrino program flagship is so called Long Baseline Neutrino Experiment or LBNE. This experiment is under development now with mass of LAr detector of ~30 kton located deep underground in South Dakota mine with distance of 1300 km from Fermilab. This experiment has over 5 sigma sensitivity to resolve neutrino mass hierarchy for any value of CP violation parameter and an excellent detector for supernova and proton decay with sensitivity better than $10^{35}$ years. Design and construction of such detector will take about a decade with data taking starting by late 2020's. Figure 4 presents overall scheme of the LBNE experiment.

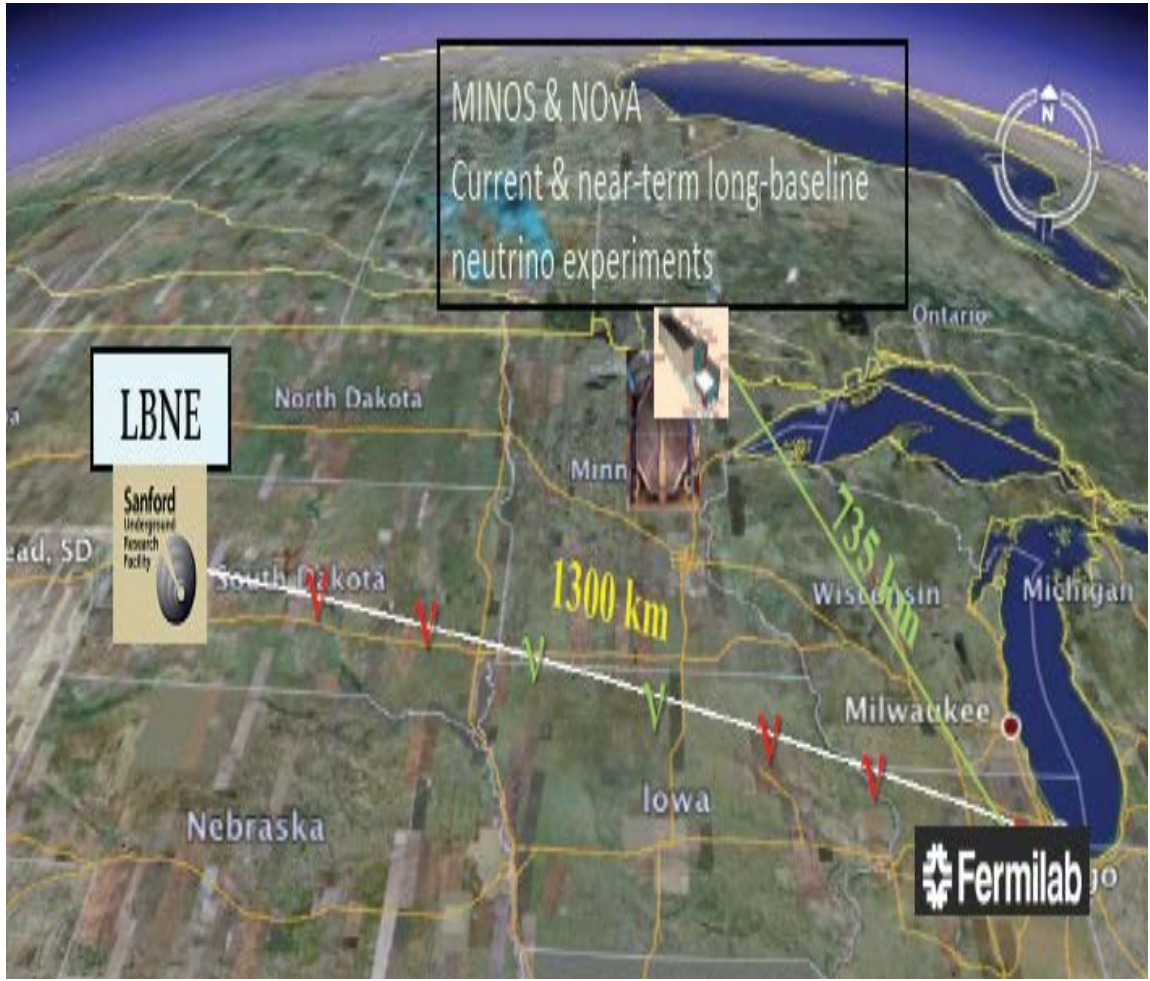

Fig. 4. LBNE experiment.

## 2.4 Muon Experiments

High intensity proton beam from Fermilab's 150 GeV accelerator of up to $2 \cdot 10^{17}$ protons per hour provides an opportunity to create very high intensity muon beams. Two experiments are planned to use such high intensity muon beams (see Fig. 2). One is g-2 experiment which is based on the BNL E-821 experiment and uses a superconducting coil moved from BNL to Fermilab. This experiment is expected to start in 2017 and have sensitivity ~4 times better than the BNL experiment. Such sensitivity is expected to address ~3 sigma difference between the g-2 experimental measurement and theoretical predictions which currently exists.

A second experiment is the Mu2e experiment which is looking for forbidden in the Standard Model decay of muon into electron after muon capture on a nuclear target. This experiment is expected to improve sensitivity to such decay by ~4 orders of magnitude (in comparison with the current limit) to $\sim 10^{-17}$. The experiment uses an intense muon beam as well as interesting idea of absorbing a large number of muons in an aluminum target.

**2.5 Colliders**

Fermilab is actively involved in the LHC program: CMS experiment and LHC upgrades. LHC experiments are critical to study all properties of the Higgs boson: mass, width, spin, couplings and others. LHC is the key machine for the next $\sim 10$ years for searches for new particles and interactions and Fermilab will be an active part of this international program.

Japan expressed strong interest in hosting the International Linear Collider (ILC) - 500 GeV $e^+e^-$ linear accelerator which will be a factory for precision studies of all known Standard Model particles as well as discovery machine for new particles and interactions. US accelerator community is capable and interested to contribute in many areas and there is strong interest in the development of detectors for the experiments at the ILC. ILC design is technically ready to go and incorporates US contributions to machine physics and technology including SCRF, high power targetry ($e^+$ source), beam delivery, damping rings, and beam dynamics.

Muons do not have high synchrotron radiation making a circular accelerator viable for multi TeV energies. The main challenge is that muons are unstable with life-time of 2 $\mu$s. The accelerator challenge is to make a large number of muons quickly and then "cool" them to focus into small diameter beams to collide. The 2x2 TeV muon collider at Fermilab layout is presented in Fig. 5. Another issue is decays and irradiation by electrons from muon decays and even neutrinos irradiation! This is an actively developing program in US with up to 10 TeV muon collider to fit on Fermilab's site.

Going to ever higher energies and smaller distances is the quest of particle physics for over 60 years. For a proton-proton collider next step is an about 100 TeV collider and active studies of options are performed at Fermilab. Figure 6 illustrates a design located in the vicinity of Fermilab with the total length of the tunnel 233 km and energy of 175 TeV. Such collider will be crucial for future progress especially if full energy LHC is not enough to reach new physics.

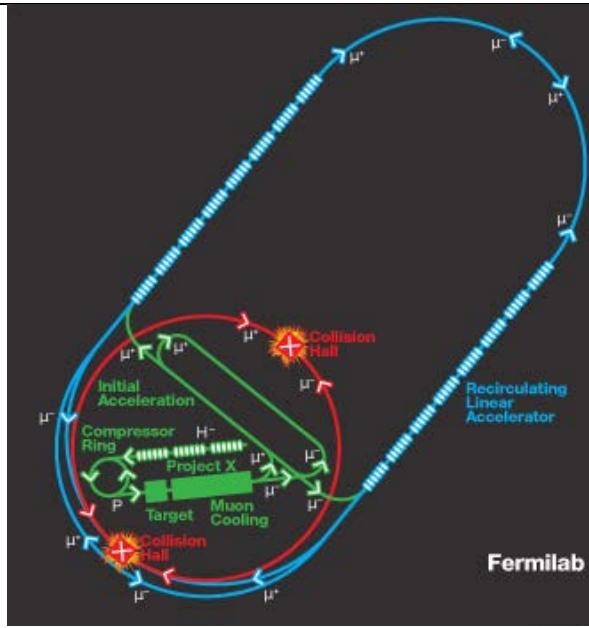

Fig. 5. Muon collider layout at Fermilab.

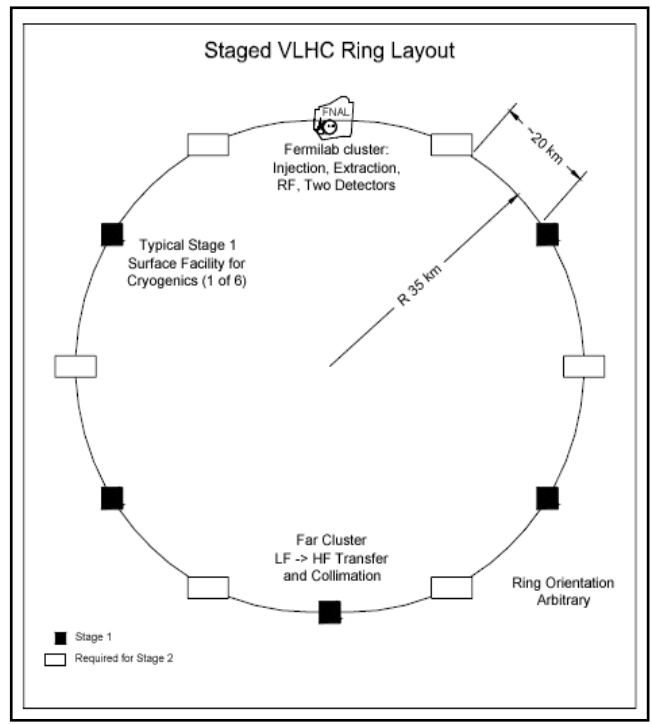

Fig. 6 175 TeV p-p collider design.

## 2.6 Dark Matter and Dark Energy

Fermilab is actively involved in studies of two astrophysics puzzles: dark matter and dark energy. In the dark matter searches key expertise of Fermilab is particle detectors. The Cold Dark Matter Search experiment or CDMS located in Sudan mine is already producing world best limits on mass and cross sections of WIMP particles. 200 kg "Super CDMS" upgrade is planned and will be located at SNOLAB. Sensitivity of Super CDMS will be below $10^{-44}$ cm$^2$ for WIMP-nucleon cross sections with a mass of ~100 GeV.

In studies of dark energy Fermilab provides a 570 megapixels silicon camera based on experience with development of silicon vertex detectors for Tevatron and LHC. The Dark Energy Survey (DES) telescope is located in Chile and started operation in September 2013. DES major scientific areas are: studies of dark matter, dark energy, supernova, solar system survey and many others.

## 3. Concluding Remarks

Particle physics in US is undergoing changes after shutdowns of the SLAC B factory and Fermilab Tevatron over last 5 years. Snowmass process created well documented list of exciting proposals for new accelerators and experiments. The most probable accelerator projects for this decade are new Fermilab linac, ILC and LHC high luminosity upgrade. Among large new upgrades/experiments planned at Fermilab are g-2, Mu2e, LBNE, ATLAS/CMS upgrades, dark matter searches. P5 process is progressing to set priorities for the coming ~10 years. There is wealth of experience at Fermilab in both accelerators and detectors and interesting program ahead.

## Acknowledgments

I would like to thank Organizers of 2014 Novosibirsk Instrumentation conference for the invitation to give this review talk and their hospitality during my stay at Novosibirsk as well as to all my Fermilab colleagues who carry out interesting program of particle physics experiments and for their help with preparations of this presentation.